\documentclass[aps,showpacs,twocolumn]{revtex4}

\usepackage{graphicx}
\usepackage{dcolumn}
\usepackage{bm}
\usepackage[squaren]{SIunits} 

\begin{document}
\title{Magnetodielectric Study in SiO$_{2}$-coated Fe$_{3}$O$_{4}$ Nanoparticle Compacts}
\author{Chung-Chieh Chang}
\author{Li Zhao}
\email{zhaoli@phys.sinica.edu.tw}
\author{Maw-Kuen Wu}
\affiliation{Institute of Physics, Academia Sinica, Taipei 11529, Taiwan}
\date{\today}

\begin{abstract}
The dielectric properties of Fe$_{3}$O$_{4}$ magnetic nanoparticles with an insulating coating layer of SiO$_{2}$ were investigated. At high temperatures, the changes of the dielectric constant and loss induced by the magnetic field are opposite in sign and strongly frequency-dependent, which originates from extrinsic magnetodielectric coupling-the Maxwell-Wagner effect combined with magnetoresistance. And the interface defects leads to the obvious hysteresis phenomena observed in the measurements. On the other hand, the strong coupling of dielectric and magnetic properties at low temperatures contradicts the Maxwell-Wagner model, suggesting the intrinsic magnetodielectric coupling. Our observations are consistent with the recent polarization switching measurements, which confirm the low-temperature multiferroic state existing in highly-lossy Fe$_{3}$O$_{4}$. And the core/shell nanostructure may provide a new route to achieve applicable magnetoelectric materials with low loss.
\end{abstract}
\maketitle

\section{INTRODUCTION}
At present, there is a worldwide revival of active interest in multiferroic materials, in which two or more ferro orders, usually magnetization and ferroelectric polarization, coexist and are coupled \cite{Fiebig}. The driving force for the multiferroic research comes not only from fundamental physics but also the potential technique application for new as magneto-electronic sensors and memory devices \cite{SWCheong, Ramesh}. Since single-phase natural multiferroic compounds are quite scarce in nature, more researcher have concentrated on developing the multi-phase multiferroic composites, which are usually composed of piezoelectric and piezomagnetic materials that can give  large magnetoelectric response via elastic coupling \cite{CWNan}. Recently, Catalan's theoretical study suggests that the magnetodielectric response can be acquired from the combination of the magnetoresistance(MR) and Maxwell-Wagner effect in heterogeneous samples and the intrinsic magnetoelectric coupling in the constituents are not necessary \cite{Catalan}. Several experimental works have been reported on phase-separating manganites \cite{Rivas} or multi-phase nanoscaled composites \cite{Lawes, Koo, Kim, Bonaedy}. The field-tunable capacitance can be achieved for possible sensor application, although the unfavorable high loss is usually be concomitant.

	Magnetite(Fe$_{3}$O$_{4}$), as the first magnetic material known by mankind, has been intensively studied for nearly one century. As an inverse spinel oxide, Fe$_{3}$O$_{4}$ has the highest known Curie temperature(860K) and good conductivity at room temperature among the ferrites, which make it appropriate for future spintronics application \cite{Coey1}. At low temperatures, Fe$_{3}$O$_{4}$ ,exhibits many exotic properties, such as the famous Verwey transition(‵125K), whose physical nature has not been fully understood till now \cite{Verwey, Walz}. At lower temperature($<$38K), early studies gave some evidence for the existence of a spontaneous polarization \cite{Rado, Kato, Miyamoto1988, Miyamoto1993}. But there exists very high loss due to the semiconducting rather insulating conductivity in Fe$_{3}$O$_{4}$, the convincing proof of ferroelectricity had been lacking due to the limitation of traditional ferroelectricity testing methods \cite{Schmid}. Very recently, the cogent ferroelectric switching in epitaxial Fe$_{3}$O$_{4}$ films is firstly demonstrated in the experiment by Alexe et al and the switchable polarization is as high as 11$\mu C$/cm2 below 20K \cite{Alexe}. The corresponding theoretical calculation further shows the ferroelectricity in low temperature phase is driven by charge ordering(CO), rather than the traditional ion displacement mechanism in classical oxide ferroelectrics \cite{Yamauchi}. Therefore, the strong multiferroicity is expected in this traditional magnetic material. The experiments on the magnetoelectric coupling are urgently needed in further studies.

In this paper, the  Fe$_{3}$O$_{4}$ nanoparticles were synthesized and further coated with the insulating SiO$_{2}$ layers, which make them suitable for practical dielectric applications. The dielectric properties in the magnetic field are thoroughly investigated. The magnetocapacitance at high temperatures are attributed to extrinsic effect.  At low temperature($<$50K), the intrinsic magnetoelectric coupling shows up consistent with appearance of ferroelectricity. The hysteresis develops as the temperature grows, which comes mainly from the carriers trapped at the defects of the interfacial layer of core/shell structure.

\section{EXPERIMENTAL}

Firstly, the hematite(Fe$_{2}$O$_{3}$) nanoparticles with elliptical shapes, were synthesized as a precursor by aging the solution of FeCl$_{3}$ and NaH$_{2}$PO$_{4}$ in a 500ml Pyrex bottle (pH=11) for 3 days. The precipitated hematite particles were centrifuged and washed several times to remove any extraneous ions. The purified hematite precipitate was filtered and dried in a vacuum oven at 100\celsius \ for 2 days. The obtained core particles were coated with uniform layers of silica by hydrolysis of tetraethyl orthosilicate(TEOS) using a procedure developed earlier \cite{Ohmori}.

The bare and SiO$_{2}$-coated hematite particles were annealed in a tube furnace in the flowing reducing gas of a CO/CO$_{2}$ 1:3 mixture at 350\celsius \ for 6h. The hematite particles were totally converted to magnetites(Fe$_{3}$O$_{4}$). The SiO$_{2}$-coated hematite nanoparticles were then pressed into pellets and subsequently sintered in pure Argon atmosphere for 24 hrs at 350\celsius. Referring to the Brinkmann's experience in annealing highly oxygen-sensitive n-type cuprates \cite{Brinkmann}, we buried the pressed pellets in the pure commercial magnetite powders to prevent the possible oxidation of Fe$^{2+}$ ion during the sintering procedure. The final products are insulating even at room temperature. For comparison, the bulk from the bare un-coated Fe$_{3}$O$_{4}$ nanoparticles was also prepared in the same procedure.

The magnetization of the samples was measured with a commercial magnetometer (SQUID VSM, Quantum Design). The magnetoresistance measurements were performed on the uncoated samples using a Quantum Design Physical Properties Measurement System(PPMS) with the standard four -probe technique. For the measurements of dielectric constant, the samples are polished to thin plates and we applied silver paint to both sides as electrodes to form a parallel plate capacitor. All our measurements were performed on the PPMS) using home-built probes. We use Cernox thermometers epoxied on the cryogenic stage very close to the sample in our set-up. We adopt the four-terminal-pair configuration in which four low-loss copper coaxial cables are used to connect the sample (glued on the cryogenic stage) to the capacitance meter. The main sources of error, residual impedance and cable length in the circuit, are carefully considered and compensated. Our measurement system was tested with the commercial standard capacitors. For magnetic field-dependent dielectric measurements at fixed temperature, the  capacitance of the samples was measured over a range of frequencies (1kHz-1MHz) while the field was swept at a low rate of 25Oe/sec. The ac excitation is 1V during the measurements.

\begin{figure}
\includegraphics[width=0.50\textwidth]{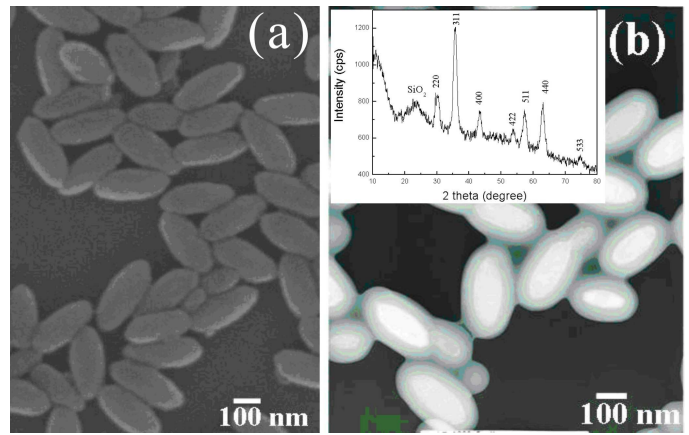}
\caption{
(a) SEM image of the uncoated Fe$_{3}$O$_{4}$ nanoparticles before coating. (b) TEM image of Fe$_{3}$O$_{4}$ nanoparticles coated with SiO$_2$ layer and the corresponding XRD pattern is shown in the upper inset.}
\end{figure}

\section{RESULTS AND DISCUSSIONS}

Both the bare and SiO$_{2}$-coated Fe$_{3}$O$_{4}$ core/shell nanoparticles have been characterized by x-ray diffraction (XRD), scanning electron microscope (SEM) and  transmission electron microscopy (TEM). The magnetite parts are ellipsoid-like with longer axis of about 240nm and shorter axis of about 100nm. The homogeneous silica coating layer is about 50nm thick, as shown in the TEM image (see Fig.1b). The corresponding XRD pattern shows no impurities such as hematite existing in our samples.

\begin{figure}
\includegraphics[width=0.50\textwidth]{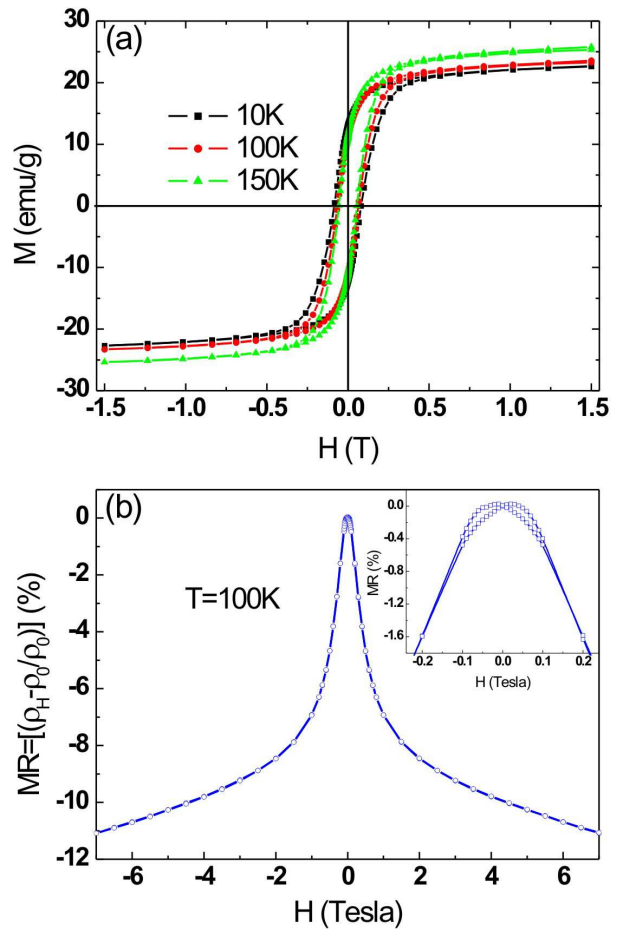}
\caption{
(a) The magnetic hysteresis loops measured at 10K, 100K and 150K. (b) MR as a function of applied field for the uncoated Fe$_{3}$O$_{4}$ nanoparticles at T=100K.}
\end{figure}

We first discuss the magnetic properties of the core/shell nanocomposites. The temperature-dependent magnetization (M-T) behaves
very similar with the bulk magnetite and no superparamagnetic behavior was observed till 300K (not shown here). Figure 2a shows the magnetic
hysteresis loops measured at different temperatures (150 K, 100 K and 10 K). The magnetic hysteresis loop of the uncoated sample
was similar, indicating the negligible influence of the SiO$_{2}$/Fe$_{3}$O$_{4}$ interface on the magnetic properties. The only
difference between these two samples is the reduced saturation magnetization (Ms) in the coated one. For example, at 10K, the Ms
of coated sample is about 23 emu/g (not shown here), only half of the uncoated one, much less than the bulk value (about 90 emu/g
for pure magnetite). The smaller magnetization is mainly due to the nonmagnetic SiO$_{2}$ and the residual surfactant,and the possible
 existence of magnetic dead layer due to a surface/interface defects layer. From the comparison the saturation magnetization
of two samples, we deduced the non-magnetic SiO$_{2}$ shells to be about 50\% of the sample, which consistent with our estimation form the TEM observation.

The field-dependent resistance (MR) is measured on uncoated Fe$_{3}$O$_{4}$ sample at T=100K (see Fig.2b). For lower temperatures,
 the resistance of the sample grows more sharply and data became more noisy as well. The accurate MR is not available. A ``butterfly"
 curve with a reversible high field slope is observed, with two maxima around the coercive field. The MR is about -11\% at H=7T. These results are coincident with experiments on the powder compacts by Coey et al, which is much higher than the intrinsic MR of magnetite obtained on the single crystals \cite{Coey1, Ziese}. The large MR is due to the field-induced alignment the magnetization of contiguous grains, which is associated with the intergranular tunneling transport of spin-polarized electrons \cite{Coey2}.  But For our SiO$_{2}$-coated samples, the kind of tunneling is prohibited even at room temperature due to thick insulating coating, which is in sharp contrast to the enhanced MR observe in Fe$_{3}$O$_{4}$ nanoparticle compact with very thin SiO$_{2}$ shells(only 3-4 nm) \cite{JShi}.

The dielectric measurement on the SiO$_{2}$-coated Fe$_{3}$O$_{4}$ core/shell nanocomposites is performed first at low temperatures. The temperature dependent dielectric constant and corresponding loss (tan$\delta$) is shown in Fig. 3. The large dispersion behaviors were observed with the characteristic of the Debye-type dipolar relaxation, which is also observed in magnetite films by Alexe et al \cite{Alexe}. This large dielectric dispersion is one common feature with the relaxor-type ferroelectric transition. The dispersion of magnetite is very similar to the CO-driven multiferroic LuFe$_{2}$O$_{4}$\cite{Ikeda}. At a given temperature T, the characteristic response frequency f (denoted as the peak in the dielectric loss) obeys a simple Anrrhenium relation,
\begin{equation}\label{formula}
f=f_{0}e^{-Q/kT}
\end{equation}
where $Q$ is the activation energy. $f_{0}$ a preconstant and the $k$ is the Boltzmann constant. Fitting out data to Eq. (\ref{formula}) gives $Q=40$ meV, the value very close to the extraction from the Alexe's data on epitaxial magnetite films (32 meV) (see the supporting online material) \cite{Alexe}.

\begin{figure}
\includegraphics[width=0.5\textwidth]{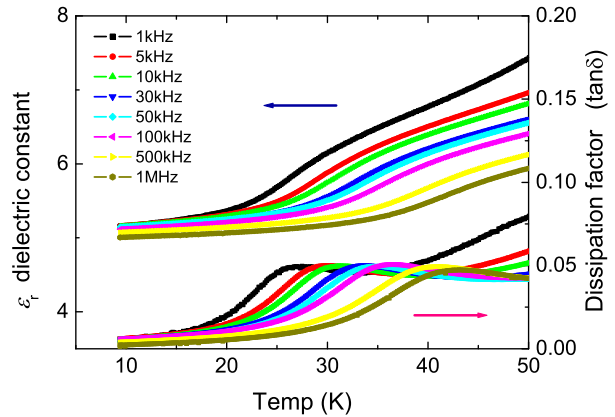}
\caption{
Temperature dependent dielectric constant and loss measured at frequencies from 1kHz to 1MHz.
}
\end{figure}

It is notable that the dielectric loss of our SiO$_{2}$-coated Fe$_{3}$O$_{4}$ core/shell composites is much less than that of pure Fe$_{3}$O$_{4}$ (usually much more than the order of 1). The very low loss makes it very suitable to perform accurate ac capacitance measurement, and also make practical device applications possible.

\begin{figure}
\includegraphics[width=0.50\textwidth]{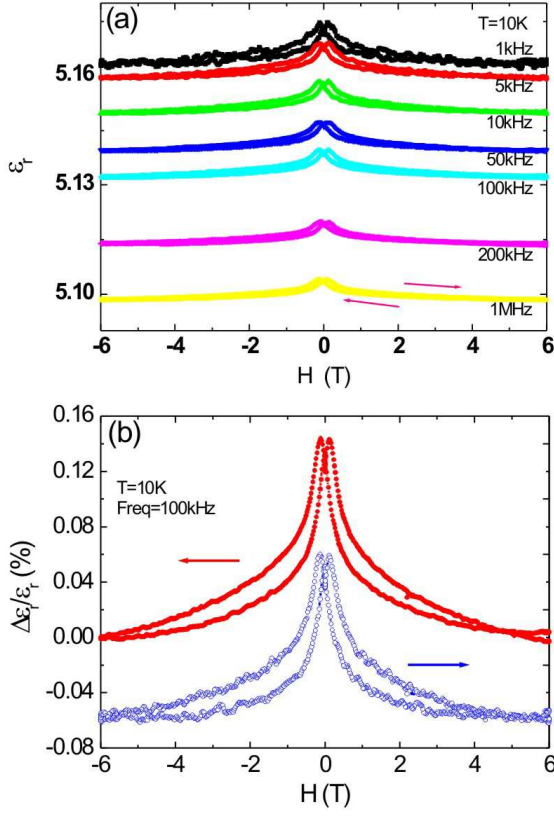}
\caption{
(a) Field dependent dielectric constant of SiO$_{2}$-coated Fe$_{3}$O$_{4}$ core/shell nanocomposite for different frequencies (1kHz to 1MHz) at T=10K. The two small arrows mark the direction of sweeping field (from 6T to -6 T then back to 6T). (b) the magnetodielectric and magnetoloss of SiO$_{2}$-coated Fe$_{3}$O$_{4}$ nanocomposite for a represent frequency(200kHz) at 10K.}
\end{figure}

The results of magnetodielectric measurements at 10K are shown in Fig.4. The magnetic field was swept from 6T to -6T (at a rate of
25Oe/sec), then returned back to 6T (denoted by the red arrows in Fig. 4a). The butterfly-shaped dielectric curves are observed for all
our probing frequencies (1 kHz-1MHz).  The dielectric constant ($\epsilon_r$) of the Fe$_{3}$O$_{4}$/SiO$_{2}$ nanocomposite shows maxima
at about ㊣0.1T, coincident with the corresponding coercivity field (see Fig. 2). As the field grows, the decrease of $\epsilon_r$ is rapid
in low field region ($|$H$|<1$T), and tends to slow down in strong field. The calculated magnetodielectric shows very weak frequency
dependence. For better clarity, the relative change of $\epsilon_r$ and dielectric loss (normalized at the value when H=-6T) at a representative
frequency (100kHz) is solely plotted in Fig.4b. The magnetodielectric reaches about 0.15\%, the same order of other magnetic
nanoparticle systems \cite{Lawes}. The corresponding loss is  about only 0.01. The magnetoloss show very similar ``butterfly" curve,
with maxima at two coercivity fields. The magnetoloss is about -6\% when $|H|=6$T.

At first glance, the similarity of magnetodielectric and MR curves (see Fig. 2) seems to suggest the extrinsic origin of observed magnetoresistance phenomena, i.e., the combination of the magnetoresistance and Maxwell-Wagner effect that has been discussed in detail by Catalan\cite{Catalan}. By our carefully consideration(as seen below), this possibility can be excluded.

According to the calculation based on Maxwell-Wagner equations, whether the MR comes from intrinsic properties of  the bulk (``core-dominated") or from intergranular spin-polarized tunneling transport (``interface-dominated"),  the decrease of $\epsilon_r$ of the composite is always accompanied by the increase of the corresponding dielectric loss, and vice versa \cite{Catalan}. A  strong frequency dependence exists in magnetodielectric and magnetoloss according to the Maxwell-Wagner equations. In our case, both magnetodielectric and magnetoloss show very similar behavior for all our testing frequencies (from 1 kHz to 1MHz). In addition, at such low temperature, the Fe$_{3}$O$_{4}$ cores become highly insulating. In our experiments, the resistance of the uncoated Fe$_{3}$O$_{4}$ nanoparticles compacts exceeds our measure limit when T$<$50K. The SiO$_{2}$-coated samples are insulating even at room temperature. The tunneling between the core/shell nanoparticles is also prohibited.  Our results suggest that the observed magnetodielectric behavior comes from the intrinsic magnetoelectric coupling of Fe$_{3}$O$_{4}$, not the extrinsic MW effect.

\begin{figure}
\includegraphics[width=0.50\textwidth]{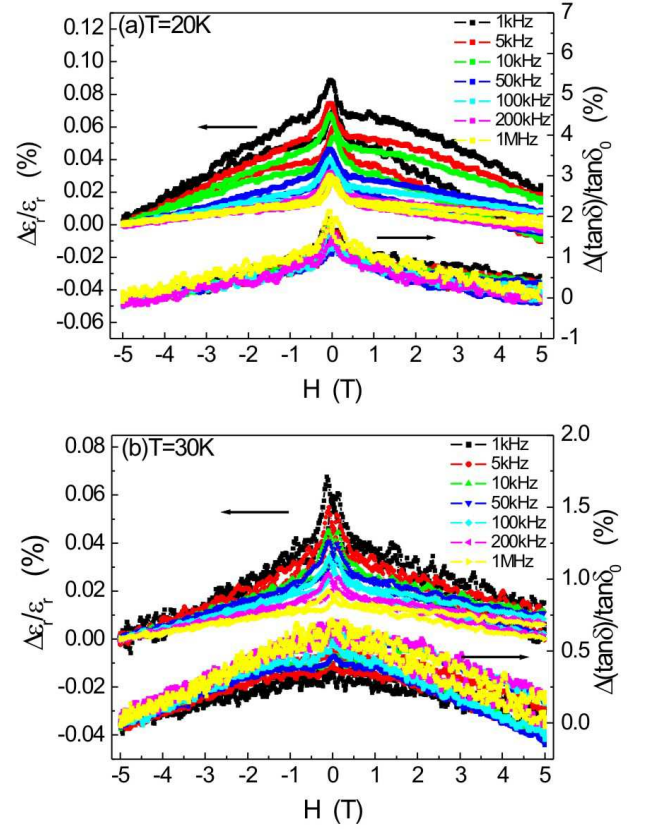}
\caption{
The magnetodielectric and magnetoloss of the SiO$_{2}$-coated Fe$_{3}$O$_{4}$ nanocomposite for different frequencies (1kHz to 1MHz) at T=20K(a) and 30K(b) respectively.}
\end{figure}

Both of the magnetodielectric and magnetoloss at higher temperatures (20K and 30K) are shown in Fig. 5. Although the data becomes very noisy, the similar but much weaker butterfly-type curves can be observed at different frequencies (1kHz to 1MHz). The sign of the magnetodielectric and magnetoloss remains the same. However their magnitudes decrease gradually as temperature increases, suggesting the magnetodielectric coupling in the Fe$_{3}$O$_{4}$ is highly temperature-dependent.

On the other hand, the frequency-dependence of the magnetodielectric and magnetoloss cannot be ignored in comparison to the 10K case. At the same time, the hysteretic behavior appears. Either magnetodielectric or magnetoloss deviate obviously from the initial value as the sweeping field returns back at +5T, which will be discussed later.

\begin{figure}
\includegraphics[width=0.50\textwidth]{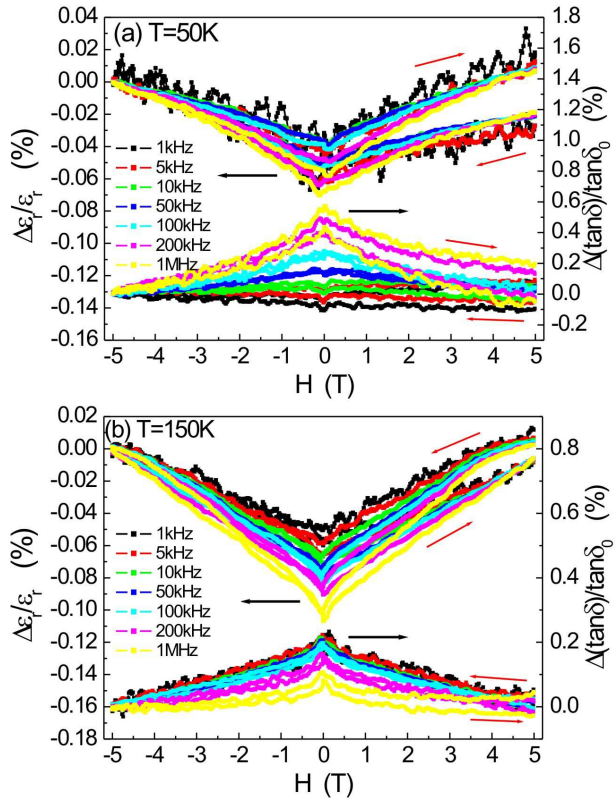}
\caption{
the magnetodielectric and magnetoloss of the SiO$_{2}$-coated Fe$_{3}$O$_{4}$ nanocomposite for different frequencies (1kHz to 1MHz) at T=50K(a) and 150K(b) respectively.  The two smaller arrows in each panel mark the direction of sweeping field (from 5T to -5 T then back to 5T).}
\end{figure}

As temperature increases further, i.e., higher than the ferroelectric transition temperature (38K, according to the experimental by Alexe \cite{Alexe}, the magnetodielectric become positive, i.e., the dielectric constant grows as $|H|$ increases. The experimental results at several typical temperatures (50K and 150K) are shown in Fig.6. At the same time, the magnetoloss remains negative. The strong frequency-dependence exists in both magnetodielectric and magnetoloss. This suggests that the intrinsic magnetoelectric coupling disappears and the extrinsic one dominates.

All these results qualitatively coincide with the theoretical prediction according the M-W equations, provided that the negative MR exists in the system from the Fe$_{3}$O$_{4}$ core. The quantitative fit to the Catalan's model is desirable to understand the microstructure and physical interactions in the nanocomposite, which is our work under way.

The hysteresis phenomena that occur at low temperatures (see Fig. 5) become more severe in our measurements at higher temperatures(see Fig. 6). The deviation from the initial value is no longer negligible and is also highly frequency-dependent. This appears to be a very slow relaxation process, which does not saturate during our experiments even the measurements were performed at the slowest field-sweeping rate (10.2Oe/sec) on PPMS. The hysteresis adds measurement noise and also leads to severe asymmetry of magnetodielectric (or magnetocapacitance in some literatures) curves, which have been observed in many previous experiments on magnetic nanoparticle system \cite{Lawes, Koo, Kim, Bonaedy}, however the corresponding discussions were absent as far as we know.

In our core/shell nanocomposites, there are be many defects primarily existing at the interfaces. The lattice misfit of the SiO$_{2}$ shell and Fe$_{3}$O$_{4}$ core, together with the low sintering temperature, leads to localized defect states of high density in the interface of core/shell or intergranular boundaries. The interface layer can be polarized when the trapped carriers tunnel through the random barriers between these states. The dielectric hysteresis may come mainly from the electric polarization of the defect states localized in the interface layer.  At very low temperatures, these trapped carriers tend to be ``freezing" and the hysteresis can be ignored. As temperature grows, the defect states are thermally activated and the contribution of the interface polarization can not be neglected. Further experimental and theoretical studies are needed.

\section{Conclusion}
In summary, we have investigated the dielectric properties of SiO$_{2}$-coated Fe$_{3}$O$_{4}$ core/shell nanocomposites. At low temperatures, the intrinsic magnetoelectric coupling is confirmed, consistent with the recent multiferroic research on Fe$_{3}$O$_{4}$ films \cite{Alexe}. At higher temperature($\geq$50K), the sign of magnetodielectric changes to positive, and comes from extrinsic magnetodielectric effect induced by the core(Fe$_{3}$O$_{4}$)-dominated MR. The hysteresis develops as the temperature grows, which possibly come from the tunneling of the trapped carries between the defects of the interfacial layer of core/shell structure. Our results emphasize the intrinsic magnetoelectric coupling at low temperatures, which is consistent with the recent confirmation of the multiferroicity in magnetite. In addition, the core/shell method provides a new way to achieve applicable magnetoelectric materials with low loss for future application.\\

\begin{acknowledgments}
The authors thank Dr Kuo-Wei Yeh and Ta-Kun Chen for their technical support in our experiments. We also thank Martin Wu for helpful discussions. LZ acknowledge the financial supportfrom the Academia Sinica and the National Science Council of Taiwan.
\end{acknowledgments}

\end{document}